\documentclass[american,
reprint,
superscriptaddress,
showpacs,
nofootinbib,
amsmath,amssymb,
aps,
pra,
floatfix,
]{revtex4-1}

\usepackage{graphicx}
\usepackage{dcolumn}
\usepackage[colorlinks=true, citecolor=blue, urlcolor=blue ]{hyperref}
\usepackage{amsmath,amssymb,amsfonts,amsthm}
\usepackage{color}
\usepackage{pxfonts,txfonts}
\usepackage{bbm} 
\usepackage{bm} 

\newcommand\numberthis{\addtocounter{equation}{1}\tag{\theequation}}

\DeclareMathOperator{\tr}{Tr}
\DeclareMathOperator{\diag}{diag}

\newcommand{\ran}{\rangle}
\newcommand{\lan}{\langle}

\newcommand{\id}{\mathbbm{1}}
\newcommand{\eeqref}{Eq.~\eqref}
\newcommand{\ccite}{Ref.~\cite}
\newcommand{\norm}[1]{\lVert #1 \rVert}

\newcommand{\N}{\mathcal{N}}

\newtheorem{theorem}{Theorem}
\newtheorem{proposition}[theorem]{Proposition}

\newtheorem{corollary}[theorem]{Corollary}

\newtheorem{conjecture}[theorem]{Conjecture}

\begin{document}


\title{Trace-distance measure of coherence}
\author{Swapan Rana}
\email{swapanqic@gmail.com} \affiliation{ICFO -- Institut de
Ci\`encies Fot`oniques, The Barcelona Institute
    of Science and Technology, 08860 Castelldefels (Barcelona), Spain}

\author{Preeti Parashar}
\affiliation{Physics and Applied Mathematics Unit, Indian Statistical Institute, 203 BT Road, Kolkata, India}

\author{Maciej Lewenstein}

\affiliation{ICFO -- Institut de Ci\`encies Fot`oniques, The
Barcelona Institute
    of Science and Technology, 08860 Castelldefels (Barcelona), Spain}
\affiliation{ICREA -- Instituci\'o Catalana de Recerca i Estudis
Avan\c cats, 08010 Barcelona, Spain}
\date{\today}

\begin{abstract}
We show that trace distance measure of coherence is a strong
monotone for all qubit and, so called,  $X$ states. An expression
for the trace distance coherence for all pure states and a semi
definite program for arbitrary states is provided. We also explore 
the relation between $l_1$-norm and relative entropy based
measures of coherence, and give a sharp inequality connecting the
two. In addition, it is shown that both $l_p$-norm- and
Schatten-$p$-norm- based measures violate the (strong) monotonicity
for all $p\in(1,\infty)$.

\end{abstract}


\maketitle


\section{\label{Sec:Int}Introduction}
It is an established fact that quantum mechanical systems differ
in many counter intuitive ways from classical systems. The figure
of merit is generally attributed to  coherence, i.e., the
possibility of quantum mechanical superpositions, which on the
level of density matrix description of quantum mechanical states
correspond to off-diagonal density matrix elements in the
computational or measurement selected basis. Many approaches have
been proposed to encompass this important feature since the
inception of quantum mechanics. Only very recently, a
resource-theoretic framework for coherence has been put forward in
Refs.~\cite{Aberg.Ar.2015,Baumgratz+2.PRL.2014}, and has been
subsequently developed \cite{Winter+Yang.Ar.2015}, and advanced
further in
Refs.~\cite{Streltsov+4.PRL.2015,Chitambar+5.Ar.2015,Streltsov+3.Ar.2015,Chitambar+Hsieh.Ar.2015}.

Quantification and interrelations between quantifiers are
important aspects in any resource theory. In general, distance
based functions are expected to be good quantifiers, subject to
the restriction imposed by the theory. In the formalism presented
in Ref.~\cite{Baumgratz+2.PRL.2014}, a non-negative convex
function $C$ defined on the  space of states $\rho$, acting on $d$
dimensional Hilbert space, is called a coherence measure if it
satisfies the following two conditions:\begin{enumerate}
    \item Monotonicity under incoherent channel $\Lambda^I$: \[C(\Lambda^I[\rho])\leq C(\rho),\]
    \item Strong monotonicity under incoherent channel $\Lambda^I$:
    \begin{equation}
    \label{Eq:Def.SM} \sum_np_nC(\rho_n)\leq C(\rho),
    \end{equation}
\end{enumerate}
    where $\rho_n:=(K_n\rho K_n^\dagger)/p_n$, $p_n:=\tr(K_n\rho K_n^\dagger)$, $K_n$'s are
    $d_n\times d$ incoherent  Kraus operators satisfying $\sum_nK_n^\dagger K_n=\id_d$. In this work we will study mainly \eeqref{Eq:Def.SM} 
    for some functions which have been proposed as possible coherence measures.

  In addition to its defining property that the resource should not increase on average under the
  \emph{free} operations, strong monotonicity has important consequences for the additivity question of convex entanglement
  measures \cite{Brandao+3.OSID.2007,Fukuda+Wolfe.JMP.2007}; additivity property in turn  simplifies further similar
  questions
  for many related information-theoretic quantities. In entanglement theory only very few measures (e.g.,
  negativity, relative entropy of entanglement, Bures' distance) are known to obey \eeqref{Eq:Def.SM}. Similarly,
  very few functions (mostly exact analogues of those entanglement measures) are known to be coherence measures.

  One of the widely used distinguishability measures, the trace distance, has been proposed as a possible candidate
  for coherence measure in \ccite{Baumgratz+2.PRL.2014}. It is formally defined as \begin{equation}
  \label{Eq:Def.Ctr}C_{\text{tr}}(\rho):=\min_{\delta\in \mathcal{I}}\norm{\rho-\delta}_1,
  \end{equation} where $\mathcal{I}$ is the set of incoherent states. The question is whether $C_{\text{tr}}(\rho)$ satisfies \eeqref{Eq:Def.SM} for all $\rho$.

  Despite its omnipresence in quantum information theory, it is not yet known whether the trace distance measure of
  entanglement $E_{\text{tr}}(\rho)$ [which is defined like \eeqref{Eq:Def.Ctr} with $\mathcal{I}$ replaced by $\mathcal{S}$,
  the set of separable states] satisfies strong monotonicity, even for the simple case of $2$-qubits. Under the extra
  assumption that the closest separable states share the same marginal states with $\rho$, $E_{\text{tr}}(\rho)$ has been
  shown to satisfy strong monotonicity \cite{Eisert+2.JPA.2003}. The difficulty of this problem is reflected by the fact
   that in general
  the closest state $\delta$ can not be determined explicitly, and even if the dimension of $\rho$ is small,
the dimension of $\rho_n$ may be arbitrarily high.
Nonetheless, as $\mathcal{I}$ has a trivial structure compared to
$\mathcal{S}$,   it seems that answering this question could be  easier for $C_{\text{tr}}(\rho)$. Here we show that $C_{\text{tr}} (\rho)$
  satisfies \eeqref{Eq:Def.SM} for all single qubit states $\rho$.

  Apart from some measures defined through the convex-roof construction \cite{Streltsov+4.PRL.2015,Qi+2.Ar.2015},
  the $l_1$-norm-based measure $C_{l_1}$ and the relative entropy based measure $C_r$ [defined later in \eeqref{Def:clp&cp}
  and \eeqref{Def:Cr}, respectively] are the only known coherence measures satisfying the strong monotonicity for all states.
  Due to its close similarity with relative entropy of entanglement $E_r$, $C_r$ has a clear physical meaning and is the
  cornerstone of resource theory of coherence \cite{Winter+Yang.Ar.2015}. In contrast,   $C_{l_1}$ has neither an exact analog
  with an entanglement measure, nor any physical interpretation yet. It is thus desirable and interesting to find any
  interrelation between them, which hopefully would give some bound on $C_{l_1}$ in terms of $C_r$.
Recently, also  the Hilbert-Schmidt distance has
  been conjectured \cite{Cheng+Hall.PRA.2015} to be a coherence measure---we show that this is not the case.

  The organization of this article is as follows:
  In Sec.~\ref{Sec:TDC} we study the properties of trace distance coherence and prove its strong monotonicity for qubits.
  We  present here also a semi
  definite program for general states. The interrelation between $C_{l_1}$ and $C_r$ is described in Sec.~\ref{Sec:l1.vs.Re}.
  It is shown that $C_{l_1}$ is an upper bound for $C_r$ for all pure states and qubit states.
  In Sec.~\ref{Sec:No.p.lp} we show that for all $p\in(1,\infty)$ neither $C_{l_p}$, nor $C_p$ satisfies (strong) monotonicity.
  We conclude with a short discussion of our results and outlook in Sec.~\ref{Sec:Discussion}.


\section{\label{Sec:TDC} Trace Distance coherence}

\subsection{Qubit and $X$-states}
To find the analytic form of trace distance coherence we have to
find the (not necessarily unique) closest incoherent state. For a
qubit the nearest incoherent state is just $\rho_{\diag}$
\cite{Shao+3.PRA.2015}. This can be seen easily: if
$\delta=\diag\{x,1-x\}$ is the nearest incoherent state to
$\rho=\binom{p~~~q}{\bar{q}~1-p}$, then $\rho-\delta$, being
Hermitian and traceless, will have eigenvalues $\pm \lambda$. So,
the required trace distance is $2\lambda$, and to minimize it we
have to minimize the determinant of $\rho-\delta$ (of course with
a negative sign, since the roots are $\pm \lambda$), which is
simply $|q|^2+(p-x)^2$, with respect to $x$. Hence the result
follows.

However, if we consider qutrits, then the expression for
eigenvalues of $\rho-\delta$, as well as the optimization becomes
very messy. As we will show later, finding an \emph{analytic form}
even for pure qutrits is almost intractable. But, we could still
find analytic expression for $C_{\text{tr}}$ (and test the strong
monotonicity) for some classes of high-dimensional states. In
doing so, we will sometimes optimize over larger 
sets of general matrices. For this purpose, we extend the definitions of $C_{\text{tr}}$ and $C_{l_1}$ to a square matrix $X$ [see also \eeqref{Def:clp&cp}], i.e.,
\begin{subequations}
	\begin{align}
C_{\text{tr}}(X)&:=\min_{D\in\Delta}\norm{X-D}_1,\\
C_{l_1}(X)&:=\min_{D\in\Delta}\norm{X-D}_{l_1}=\sum_{i\neq j}|X_{ij}|,
	\end{align}
\end{subequations}
$\Delta$ being the set of diagonal matrices. The following result will be used extensively.
\begin{proposition}
    \label{Prop:Any.2x2.diag} Let $A$ be a $2\times 2$ matrix with complex entries
    and $D$ be its closest diagonal matrix in trace norm. Then $D=\diag(A)$ and hence $C_{\text{tr}}(A)=C_{l_1}(A)$.
\end{proposition}
Note that the trace norm of $A-D$ is non-negative and always
bounded in finite dimension, e.g., by sum of individual trace
norms. Hence, we do not require any further restriction (like
normality, positivity) on $A,D$.

As usual, let $A=\binom{a~b}{c~d}$, $D=\binom{x~0}{0~y}$. If the
singular values of $A-D$ are $\sigma_1$ and $\sigma_2$, then
    \begin{align*}&\norm{A-D}_1^2=(\sigma_1+\sigma_2)^2\\
    &=\tr\left[(A-D)(A-D)^\dagger\right]+2\left|\det(A-D)\right|\\
    &=|a-x|^2+|d-y|^2+|b|^2+|c|^2+2\left|(a-x)(d-y)-bc\right|\\
    &\geq|a-x|^2+|d-y|^2+|b|^2+|c|^2+2\left||(a-x)||(d-y)|-|b||c|\right|\\
    &\geq \left(|b|+|c|\right)^2,\numberthis
    \end{align*}
with equality iff $D=\diag(A)$. In the last inequality, we have
used the fact that $p^2+q^2+2|pq-r|\geq 2|r|$, and equality holds
iff $p=q=0$.\qed

The purpose of considering general matrices instead of states is to get
rid of the positivity and the unit trace conditions. Thus we are minimizing over a larger set.
This proposition immediately implies the following facts:
\begin{corollary}
    \label{Cor:Ctr(Qubit.Oplus.Scalar)}
    Let $A_i$ be $2\times 2$ complex matrices and $x_i$, $y_j$ be complex numbers. Then \begin{equation}\label{Eq:Ctr(Qubit.Oplus.Scalar)}
    C_{\text{tr}}\left(\mathop{\oplus}_ix_iA_i\mathop{\oplus}_{j}y_j\right)=\sum_i|x_i|C_{\text{tr}}(A_i)=C_{l_1}\left(\mathop{\oplus}_ix_iA_i\mathop{\oplus}_{j}y_j\right).
    \end{equation}
\end{corollary}
This improves the theorem of \ccite{Shao+3.PRA.2015}, in the sense that it is readily applicable to direct sum of qubits. States of the form $x\oplus A$ were considered therein.
\begin{corollary}
    \label{Cor: Monotonicity.Ok.Qubit.Trace.Distance}
    The strong monotonicity is satisfied by $C_{\text{tr}}(\rho)$ for any $2\times 2$ matrix $\rho$.
\end{corollary}
In \ccite{Shao+3.PRA.2015}, the authors showed that if the dimensions of the Kraus operators are restricted to three, then the strong monotonicity is satisfied by all qubit states. We show here that this is always true irrespective of the dimensions of the Kraus operators involved.

Using the fact that $\norm{X}_p\leq\norm{X}_{l_p}$ for any
matrix $X$ and $1\leq p\leq 2$ \cite[pp.~50]{Zhan.S.2002}, we
have
    \begin{align*}
    C_{\text{tr}}(\rho_n)&=\norm{\rho_n-\delta_n^\star}_1\\
    &\leq \norm{\rho_n-\diag(\rho_n)}_1\\
    &\leq \norm{\rho_n-\diag(\rho_n)}_{l_1}=C_{l_1}(\rho_n).\numberthis
    \end{align*}

So multiplying by $p_n$, summing over $n$, using the fact that
$C_{l_1}$ satisfies the strong monotonicity condition
\cite{Baumgratz+2.PRL.2014}, and finally
$C_{l_1}(\rho)=C_{\text{tr}}(\rho)$,  proves the result. By the
same reasoning, strong monotonicity is satisfied for all matrices
$A$ with $C_{\text{tr}}(A)=C_{l_1}(A)$, in particular the matrices
in \eeqref{Eq:Ctr(Qubit.Oplus.Scalar)}.\qed

Let us now mention an interesting class of states, the so called
$X$\emph{-states}, albeit we do not assume anything (not even
normality) except its shape, for which $C_{\text{tr}}$ has an
analytic expression, also satisfies strong monotonicity.

\begin{proposition}
    Let $X$ be an $n\times n$ complex matrix  with non-zero elements only along
    its diagonal and anti-diagonal, $x_{ij}=0$ for $j\neq i,n+1-i$. The nearest diagonal matrix to $X$ in
    trace norm is given by $\diag(X)$. Therefore $C_{\text{tr}}(X)=C_{l_1}(X)$ and hence $C_{\text{tr}}(X)$ satisfies
    strong monotonicity.
\end{proposition}
While calculating trace norm, the matrix $X-\delta$ is a special
class of the matrices appearing in
\eqref{Eq:Ctr(Qubit.Oplus.Scalar)}, and hence the result follows
from Cor.~\ref{Cor:Ctr(Qubit.Oplus.Scalar)} and Cor.~\ref{Cor:
Monotonicity.Ok.Qubit.Trace.Distance}.\qed

We should mention that calculation of trace distance coherence for
a very specific class of $X$-states (with only three real parameters)
has been considered in recent literature
\cite{Bromley+2.PRL.2015,Cianciaruso+2.Ar.2015}.

\subsection{Pure states}

Finding the closest incoherent state becomes intractable just
beyond qubits. For a pure state $|\psi\ran$, the intuitively expected nearest
incoherent state is  $\delta=\diag\{|\psi\ran\lan\psi|\}$.
Unfortunately, this is not necessarily true for dimension higher
than 2. As an example, for
$|\psi\ran=2/3|0\ran+2/3|1\ran+1/3|2\ran$,  $\diag\{1/2,1/2,0\}$
is closer than $\diag(|\psi\ran\lan\psi|)$. We will now show why
it is difficult to have an analytic formula, even for the simple
case of pure qutrits.

 Let $|\psi\ran=\sum_i\sqrt{\lambda_i}|i\ran$ be given (if required, we remove any phase by a diagonal unitary, which is an incoherent operation) and let $\delta=\sum_i \delta_i|i\ran\lan i|$ be its nearest diagonal state.  Then by Weyl's inequality \cite[pp.~62]{Bhatia.S.1997} $\lambda_{i+j-1}^{\downarrow}(A-B)\leq \lambda_{i}^{\downarrow}(A)-\lambda_{n-j+1}^{\downarrow}(B)$,
 the matrix $H=|\psi\ran\lan\psi|-\delta$ has exactly one positive eigenvalue. Let it be $\alpha$. Since $H$ is traceless, the sum of the rest of its eigenvalues must be $-\alpha$, and hence $\|H\|_1=2\alpha$. The problem is thus to find the maximum (as only one is positive) eigenvalue of $H$ and minimize it with respect to $\delta_i$'s.

 As usual, we have to solve the characteristic equation for $H$, namely, $\det(xI-H)=0$. So, let us first calculate the determinant (see \cite{Anderson.LAA.1996} for more general case). Writing \[ xI-H=\begin{pmatrix}x+\delta_1&&&\\&x+\delta_2&&\\&&\ddots&\\&&&x+\delta_d
 \end{pmatrix}-|\psi\ran\lan\psi|,\] we use the Sherman-Morrison-Woodbury formula for
 determinants \cite[pp.~19]{Horn+Johnson.CUP.2012}:\[\det(\mathbf A+\mathbf u\mathbf v^t)=(1+\mathbf v^t\mathbf A^{-1}\mathbf u)\det\mathbf A.\] Therefore
 the required determinant is
 \[\det(xI-H)=\left[1-\sum\limits_{i=1}^d\frac{\lambda_i}{x+\delta_i}\right]\prod\limits_{i=1}^d(x+\delta_i).\]
For positive roots, the out-most factors are nonzero and we obtain
the equation \begin{equation}
 \label{Eq: Main.constraint.charcetristic.pure}\sum\limits_{i=1}^d\frac{\lambda_i}{x+\delta_i}=1.
 \end{equation}
 \eeqref{Eq: Main.constraint.charcetristic.pure} can be viewed as a (monic) polynomial equation in $x$ of degree $d$.
 We have to find its largest root (all roots are real)  and then minimize that with respect to $\delta_i$.
 Unless $d=2$ (where the roots are of the form $(b\pm \sqrt{b^2-4c})/2$, thereby the largest root is the one with the $+$ sign), there is no
 simple  way to characterize the largest root $x^\star$, and hence in general, no simple way to get a general explicit expression
 for $C_\text{tr}$. Note also that we can consider $\lambda_i\neq 0$, if some $\lambda_i=0$, then the problem is reduced to the case with $d=\#\left\{\lambda_i\neq 0\right\}$

 Nonetheless, the above analysis is quite useful since we have \begin{equation}
 \label{Eq:Analytic.TDC.Pure} C_{\text{tr}}(|\psi\ran)=2 \min_{\delta_i\geq 0,\,\sum\delta_i=1}\;\max_{x\geq 0}\left\{\sum\limits_{i=1}^d\frac{\lambda_i}{x+\delta_i}=1\right\}.
 \end{equation}
 The right hand side (RHS) of \eeqref{Eq:Analytic.TDC.Pure} can be written as the following
  optimization problem:
 
    \begin{align*}
    \text{ Minimize } &~\frac{2}{d} \left(\sum\limits_{i=1}^d\frac{\lambda_i}{\delta_i}-1\right)\\
    \text{subject to  }&\left\{\begin{array}{rl}
  \frac{\lambda_i}{\delta_i}&\geq \frac{1}{d}\left(\sum\limits_{i=1}^d\frac{\lambda_i}{\delta_i}-1\right),\,\forall\, i=1,2,\dotsc,d,\\
    \sum\limits_{i=1}^d \delta_i&\leq 1,\numberthis\label{Eq:optimization.pure}\\
    \delta_i&\geq 0,\,\forall\,  i=1,2,\dotsc,d.
    \end{array}\right.
    \end{align*}
    
To see this equivalence, first note that \eeqref{Eq:Analytic.TDC.Pure} could be rewritten as \begin{equation}
\label{Eq:equiv.optimization.pure.0} C_{\text{tr}}(|\psi\ran)=2 \min_{\delta_i\geq 0,\,\sum\delta_i=1,\,x>0}\;\left\{\sum\limits_{i=1}^d\frac{\lambda_i}{x+\delta_i}\leq 1\right\}. 
\end{equation}
Let $x^{\star}>0$ and $\delta^{\star}=(\delta_1^{\star},\delta_2^{\star},\dotsc,\delta_d^{\star})$ be the optimal value of $x,$ and $\delta$ in the RHS of \eqref{Eq:equiv.optimization.pure.0}.
Define \begin{subequations}
	\begin{align}
	\delta_i&:=\frac{\lambda_i}{(x^{\star}+\delta^{\star}_i)},\quad i=1,2,\dotsc,d,\label{Eq:def.deltaj.1}\\
	\Rightarrow\quad& (x^{\star}+\delta^{\star}_i)=\frac{\lambda_i}{\delta_i},\quad i=1,2,\dotsc,d.\label{Eq:def.deltaj.2}
	\end{align}
\end{subequations}
Summing \eeqref{Eq:def.deltaj.2} and using $\sum\delta_i^{\star}=1$, we have
\[x^{\star}=\frac{1}{d} \left(\sum\limits_{i=1}^d\frac{\lambda_i}{\delta_i}-1\right), \] and hence from \eeqref{Eq:def.deltaj.2} and \eeqref{Eq:def.deltaj.1},
\[\frac{\lambda_i}{\delta_i}\geq x^{\star}=\frac{1}{d} \left(\sum\limits_{i=1}^d\frac{\lambda_i}{\delta_i}-1\right)\,(>0),\quad i=1,2,\dotsc,d. \]

Thus the solution of \eeqref{Eq:equiv.optimization.pure.0} corresponds to the solution of \eeqref{Eq:optimization.pure}. Conversely, if $\delta^{\star}=(\delta_1^{\star},\delta_2^{\star},\dotsc,\delta_d^{\star})$ is the optimal values of $\delta$ in \eeqref{Eq:optimization.pure}, then one easily verifies that \[x^{\star}:=\frac{1}{d} \left(\sum\limits_{i=1}^d\frac{\lambda_i}{\delta_i^{\star}}-1\right),\,\delta_i:=\frac{\lambda_i}{\delta_i^{\star}}-x^{\star},\quad i=1,2,\dotsc,d\] correspond to the optimal $x$ and $\delta$ in \eeqref{Eq:equiv.optimization.pure.0}.

\subsection{Arbitrary states}
In contrast to trace distance entanglement, we could formulate
a semi-definite program to calculate 
$C_{\text{tr}}(\rho)$ for any arbitrary state $\rho$. The main
idea is that any Hermitian matrix $\rho$ can be written as a
difference of two positive semi definite matrices,
$\rho=\rho^+-\rho^-$, with $\rho^{\pm}\geq 0$. Then
$\norm{\rho}_1$ is Tr$(\rho^++\rho^-)$ minimized over all such
decompositions of $\rho$. Thus $C_{\text{tr}}(\rho)$ is the optimal
value of the following semi definite problem (SDP)
\cite{A.Winter.is.Great}:
     \begin{align*}
    \text{ Minimize } &~\tr(P+N)\\
    \text{subject to  }&\left\{\begin{array}{rl}
    P-N &=\rho-\delta,\\
        \tr\delta&= 1,\\
    \delta&\text{ is diagonal,}\numberthis\\
    P,N,\delta&\geq 0.
    \end{array}\right.
    \end{align*}

We have used this SDP to check the strong monotonicity for random states (however, we were not
able to generate the incoherent channels uniformly). Despite our numerical and analytic attempts, no examples violating strong monotonicity were found. This leads us to conjecture that strong monotonicity of $C_{\text{tr}}$ is satisfied by all states. 


\section{\label{Sec:l1.vs.Re}Relation between $C_{l_1}$ and $C_r$}
Analogously to the relative entropy of entanglement, the relative
entropy of coherence is defined \cite{Baumgratz+2.PRL.2014} as
\begin{equation}\label{Def:Cr}
C_r(\rho):=\min_{\delta\in\mathcal{I}}S(\rho\|\delta).
\end{equation}  The minimization could be solved analytically \cite{Baumgratz+2.PRL.2014}, leading to
$C_r(\rho)=S(\rho_{\diag})-S(\rho)$,  where
$S(\rho)=-\tr(\rho\log_2\rho)$ is the von Neuman entropy. Note that for pure states $C_{l_1}$ is somewhat like
the negativity $\N$ \cite{Vidal+Werner.PRA.2002} of a bipartite
state,
\[C_{l_1}\left(|\psi\ran:=\sum\sqrt{\lambda_i}|i\ran\right)=\left(\sum
\sqrt{\lambda_i}\right)^2-1=2\N\left(|\phi\ran:=\sum\sqrt{\lambda_i}|ii\ran\right).\]
In entanglement theory, relations between $E_r$ and $\N$ have
been studied extensively (albeit mainly for two qubits, see, e.g.,
\cite{Verstraete+3.JPA.2001,Miranowicz+Grudka.JOB.2004,Miranowicz+3.PRA.2008}).
The aim of this section is to derive interrelations between $C_r$ and
$C_{l_1}$.

\subsection{Pure states}

For pure qubit states, the relation is $C_{l_1}\geq C_r$, which is
exactly the well known upper bound for the binary entropy function
\[2\sqrt{x(1-x)}\geq H_b(x):=-x\log_2 x-(1-x)\log_2 (1-x).\] For
higher dimensional pure states, we will exploit two known results --- one from entanglement theory and the other from information theory.
 It is well known
\cite{Vidal+Werner.PRA.2002} that the logarithmic negativity
$E_{\N}:=\log_2(1+2\N)$ is an upper bound on distillable entanglement
which coincides with $E_r$ for pure states,
    \begin{align*}
    \log_2\left(1+2\N(|\phi\ran)\right)&\geq E_r(|\phi\ran)\\
    \Rightarrow\,\log_2\left(1+C_{l_1}(|\psi\ran)\right)&\geq C_r(|\psi\ran)\\
    \Rightarrow\, C_{l_1}(|\psi\ran)&\geq 2^{C_r(|\psi\ran)}-1.\numberthis\label{Eq:Cl1.geq.re.pure}
    \end{align*}
Note that this bound is tight in the sense that equality holds for
maximally coherent states in any dimension. 

There is another simple inequality between $C_{l_1}$ and $C_r$, namely $C_{l_1}\geq C_r$ for all pure states.  Although generally not sharp, this inequality is independent from that in \eeqref{Eq:Cl1.geq.re.pure}.  To prove it, note that it follows \cite{Lin.IEEE.1991} from the recursive property of entropy function,
\begin{align*}
\frac{1}{2}H(\lambda)&\leq \sum_{i=1}^{d-1}\sqrt{\lambda_i\sum_{j=i+1}^d\lambda_j}\leq \sum_{i=1}^{d-1}\sqrt{\lambda_i}\left(\sum_{j=i+1}^d\sqrt{\lambda_j}\right),\\
\Rightarrow\,C_r(|\psi\ran)&\leq C_{l_1}(|\psi\ran).\numberthis\label{Eq:Cl1.geq.Cre.pure}
\end{align*}
Combining Eqs.~\eqref{Eq:Cl1.geq.re.pure}-\eqref{Eq:Cl1.geq.Cre.pure} we have the following result.

\begin{proposition}
    \label{Prop:l1.geq.re.for.pure.ln} For all pure states $|\psi\ran$, \begin{equation}
    \label{Eq:cl1.geq.re.psi} C_{l_1}(|\psi\ran)\geq\max\left\{C_r(|\psi\ran),\; 2^{C_r(|\psi\ran)}-1\right\}.\end{equation}
\end{proposition}

\begin{figure}[h]
    \begin{center}
        \includegraphics[height=4.3cm]{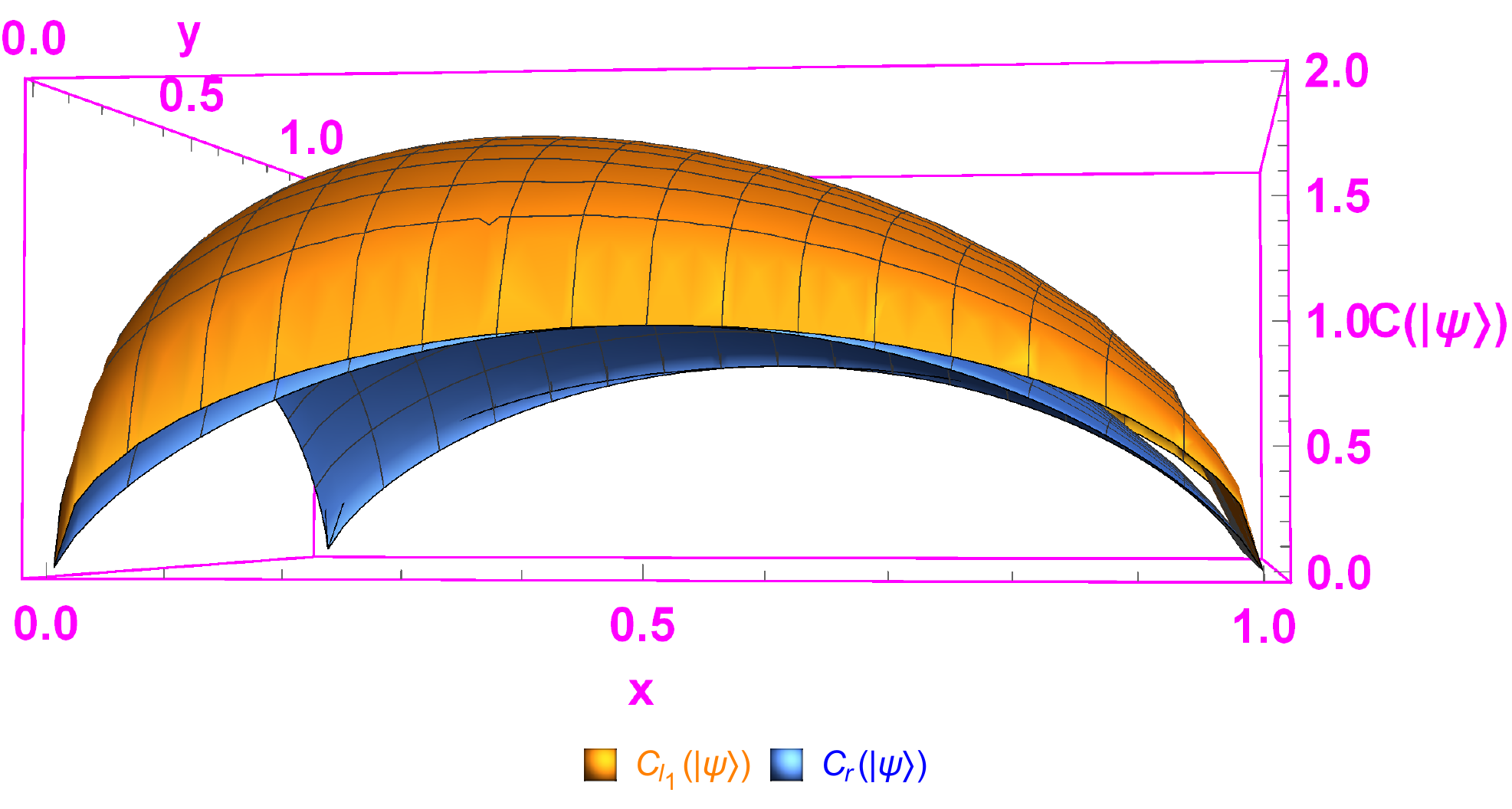}\caption{(Color online) $C_{l_1}(|\psi\ran)$ and $C_r(|\psi\ran)$ for general qutrit $|\psi\ran$.} 
        \label{Fig:cl1.geq.er.d=3}
    \end{center}
\end{figure}

The variation of these bounds could be visualized for arbitrary qutrits. In this case, the $\lambda_i$'s can be taken as $x,(1-x)y,(1-x)(1-y)$ and Fig.~\ref{Fig:cl1.geq.er.d=3} shows the plot of $C_{l_1}$ and $C_r$ as a function of $x,y\in[0,1]^2$. Note that $C_{l_1}\geq C_r$ gives independent bound than that of \eeqref{Eq:Cl1.geq.re.pure}. For example, let $x=1/500$, $y=1/5$, then $C_{l_1}(|\psi\ran)=0.9182$, $C_r=0.7413$, while the bound in \eeqref{Eq:Cl1.geq.re.pure} gives $C_{l_1}\geq 0.6717$. On the other hand \eeqref{Eq:Cl1.geq.re.pure} gives equality for all maximally coherent states.

Note also that, \eeqref{Eq:cl1.geq.re.psi} improves the known bound on distillable entanglement (in terms of logarithmic negativity), for all pure states.

\subsection{Arbitrary states}
As usual, finding a better bound is more difficult for mixed states. Since the $C_{l_1}$ measure does not have any role in entanglement theory so far, we could use the inequality $C_{l_1}\geq C_{\text{tr}}$, resulting in some rough bounds, due to the proportionality constant already introduced in this step. It is then tempting to use Fannes's inequality\cite{Fannes.CMP.1973}, but unfortunately it gives nothing useful:
\begin{align*}
C_r(\rho)&=S\left(\rho\|\rho_{\diag}\right)\\
&\leq \norm{\rho-\rho_{\diag}}_1\log_2 d+\frac{1}{e\ln 2}\\
&\leq C_{l_1}(\rho)\log_2 d + \frac{1}{e\ln 2}.\numberthis\label{Eq:Fannes.use}
\end{align*}
The relation between $C_{\text{tr}}$ and $C_r$ could be
drastically sharpened [than the one mentioned in
\eeqref{Eq:Fannes.use}] using Fannes-Audenaert bound
\cite{Audenaert.JPA.2007}; unfortunately this bound is not
monotonic in $C_{\text{tr}}$ and hence not applicable to
$C_{l_1}$.

It turns out that we can use an inequality between Holevo information $\chi$ and trace norm. For an ensemble $\mathcal{E}:=\{p_i,\rho_i\}$, the Holevo information is defined as \[\chi(\mathcal{E}):=S\left(\sum_ip_i\rho_i\right)-\sum_ip_iS(\rho_i),\] and it satisfies \cite{Audenaert.JMP.2014} \begin{equation}
\label{Eq:Holevo.chi.leq.TN} \chi(\mathcal{E})\leq H(p) t,\quad t:=\max_{i,j}\norm{\rho_i-\rho_j}_1/2.
\end{equation}
The next ingredient we will use is the fact that for any square
matrix $X$, there are sets of diagonal unitary matrices $\{U_k\}$
such that \begin{equation}
\label{Eq:Diag.rho.as.ensemble}\diag(X)=\frac{1}{r}\sum_{k=0}^{r-1}U_kXU_k^\dagger.
\end{equation}
At least two such sets of unitaries are known \cite{Bhatia.AMM.2000}, one with $r=d=$ order of $X$, and $U_k=U^k$, $U=\diag\{1,\omega,\omega^2,\dotsc,\omega^{d-1}\}$, $\omega:=e^{2\pi i/d}$. The other one is with $r=2^d$ and $U_k$'s are $\diag\{\pm 1,\pm 1,\dotsc,\pm 1 \}$. Since the second choice involves more terms, in general it leads to inferior bounds. Employing these tools, it follows that
    \begin{align*}
    C_r(\rho)&=S[\diag(\rho)]-S(\rho)\\
    &=\chi\left(\left\{\frac{1}{d},U_k\rho U_k^\dagger\right\}\right)\\
    &\leq t \log_2 d. \numberthis\label{Eq:CR.leq.tlgd}
    \end{align*}
Now let the maximum in the definition of $t$ occur for the pair
$\rho_i,\rho_j$. Since $\{U_k\}$ forms a multiplicative group, we
will have $U_i^\dagger.U_j=U_k$ for some $k$. Then \begin{align*}
2t&=\norm{U_i\rho U_i^\dagger-U_j\rho U_j^\dagger}_1=\norm{\rho-U_k\rho U_k^\dagger}_1\\
&\leq\norm{\rho-U_k\rho U_k^\dagger}_{l_1}\\
&\leq2 C_{l_1}(\rho).
\end{align*}
Plugging into \eeqref{Eq:CR.leq.tlgd}, we get the following result.
\begin{proposition}
    For any $d$-dimensional state $\rho$, \begin{equation}
    C_r(\rho)\leq\log_2 d\;\,C_{l_1}(\rho).
    \end{equation}
\end{proposition}
Note that for qubits it is already sharp, coinciding with the
bound for pure states. Our numerical study suggests that the
inequality could be sharpened to just $C_{l_1}\geq C_r$, but we
could not manage to get rid of this rather annoying multiplicative
factor. We thus make the following conjecture.
\begin{conjecture}
	\label{Conj:cl1.geq.max} For all states $\rho$, \[C_{l_1}(\rho)\geq C_r(\rho).\]
\end{conjecture}

\section{\label{Sec:No.p.lp} All other $l_p$-norm and Schatten-$p$-norm}
For an $m\times n$ matrix $X=(x_{ij})$ and $p\in[1,\infty)$, the $l_p$-norm and Schatten-$p$ norms are usually defined as
    \begin{align*}
    \norm{X}_{l_p}&:=\left(\sum_{i,j}|x_{ij}|^p\right)^{1/p},\\
    \norm{X}_p&:=\left(\tr|X|^p\right)^{1/p}=\left(\sum_{i}^r\sigma_i^p\right)^{1/p},
    \end{align*}
where $\sigma_i$'s are the non-zero singular values of $X$, i.e., eigenvalues of $|X|:=\sqrt{X^\dagger X}$, and $r$ is the rank of $X$. The coherence measure based on the distance induced by these norms are defined as
\begin{subequations}\label{Def:clp&cp}
    \begin{align}
    C_{l_p}(\rho)&:=\min_{\delta\in\mathcal{I}}\norm{\rho-\delta}_{l_p},\\
    C_{p}(\rho)&:=\min_{\delta\in\mathcal{I}}\norm{\rho-\delta}_{p}.
    \end{align}
\end{subequations}

In \ccite{Baumgratz+2.PRL.2014}, the authors have shown that
$C_{l_1}$ satisfies strong monotonicity (and $C_1=C_{\text{tr}}$
is the subject of this paper). They have also considered coherence
measure based on the distance induced by the \emph{square} of
$l_2$-norm and gave an example to show that it does not satisfy
strong monotonicity.  Although a coherence measure need not be
induced by a norm (e.g., $C_r$ is based on relative entropy which
is neither a distance for being asymmetric in its arguments, nor a
metric for violating triangular inequality), the counterexample
provided in \ccite{Baumgratz+2.PRL.2014} does not violate strong
monotonicity if we take just the $l_2$-norm, instead of its
square. Based on this observation it has been conjectured in
\ccite{Cheng+Hall.PRA.2015} that $l_2$-norm induces a legitimate
coherence measure. In this section we will show that it is not the
case. We will prove the following result:

\begin{proposition}\label{Prop:Clp&Cp.not.SM}
    For all $p\in(1,\infty)$, there are states violating strong monotonicity for both the measures $C_{l_p}$ and $C_p$,
    thereby neither is a good measure of coherence.
\end{proposition}

Before presenting our counterexample, let us mention that
$\norm{.}_2^2$ (in general $\norm{.}_{l_p}^p$ for $1<p<\infty$)
need not be a norm, as it does not satisfy the triangular
inequality \[\|a+b\|_{l_2}^2\leq\|a\|_{l_2}^2+\|b\|_{l_2}^2.\]
(It is not necessarily true when $a,b$ are tensors,
matrices, vectors, complex numbers, or even real numbers). The
\emph{homogeneity condition} of a norm is violated by
$\|.\|_{l_2}^2$. This is the reason for the apparent violation of
monotonicity by the counterexample provided in
\ccite{Baumgratz+2.PRL.2014}. Indeed the combined state and channel provided in the
example satisfies strong monotonicity inequality for any $l_p$-norm. In
particular, with those $\{K_n\}$, all states (qutrit, for the
dimensions of $K_n$'s) satisfies the strong monotonicity in $l_2$-norm,
as
\begin{subequations}
    \begin{align*}
    \sum_{i=1}^2p_iC_{l_2}(\rho_i)&=p_2C_{l_2}(\rho_2)\\
    &=\sqrt{2} \left(|\beta||c|+|\alpha||e|\right)\\
    &\leq \sqrt{2} \sqrt{|b|^2+|c|^2+|e|^2}\\
    &=C_{l_2}\left(\rho:=\begin{bmatrix}
    a&b&c\\\bar{b}&q&e\\\bar{c}&\bar{e}&f
    \end{bmatrix}\right).
    \end{align*}
\end{subequations}

However, with this judicious choice of $\{K_n\}$ with
$\alpha=\beta$, and $\rho$ with $b=0$, $c=e$, the strong monotonicity
inequality for $l_p$ norm becomes $2^p\leq2^2$, which is violated
by all $p\in(2,\infty)$.

It is well known \cite{Ozawa.PLA.2000} that the distance induced
by $l_2$-norm (see also \cite{Garcia+3.JMP.2006} for Schatten-$p$ norms) is not contractive under CPTP maps. Since a
coherence measure has to be contractive under (incoherent) CPTP
maps, there is no reason to think of $C_{l_2}$ to be a good measure
of coherence. To end this discussion, we give the following
counterexample:
\[\rho =\frac{1}{4}\begin{pmatrix}
1 & 0 & a & 0 \\
0 & 1 & 0 & b \\
\bar{a} & 0 & 1 & 0 \\
0 & \bar{b} & 0 & 1 \\
\end{pmatrix}, K_1=\begin{pmatrix}
0 & 0 & 0 & 0 \\
1 & 0 & 0 & 0 \\
0 & 0 & 0 & 0 \\
0 & 0 & 1 & 0 \\
\end{pmatrix},
K_2=\begin{pmatrix}
0 & 1 & 0 & 0 \\
0 & 0 & 0 & 0 \\
0 & 0 & 0 & 1 \\
0 & 0 & 0 & 0 \\
\end{pmatrix}.\]
Then \[\rho_1=\frac{1}{p_1}
\begin{pmatrix}
0 & 0 & 0 & 0 \\
0 & \frac{1}{4} & 0 & \frac{a}{4} \\
0 & 0 & 0 & 0 \\
0 & \frac{\bar{a}}{4} & 0 & \frac{1}{4} \\
\end{pmatrix},\, \rho_2=\frac{1}{p_2}
\begin{pmatrix}
\frac{1}{4} & 0 & \frac{b}{4} & 0 \\
0 & 0 & 0 & 0 \\
\frac{\bar{b}}{4} & 0 & \frac{1}{4} & 0 \\
0 & 0 & 0 & 0 \\
\end{pmatrix},\, p_1=p_2=\frac{1}{2}.
\]
In order for $\rho$ to be a state, we must have $|a|,|b|\leq 1$. The
strong monotonicity for the $C_{l_p}$ measure reads
\[\left(|a|+|b|\right)^p\leq |a|^p+|b|^p,\]which is violated
\cite{Explain.1} by the entire class with $ab\ne 0$ and for all
$p\in(1,\infty)$.

Now we move to the calculation for $C_p$. It turns out that we do
not need to calculate anything further. Note that if we assume that the
matrices in Proposition~\ref{Prop:Any.2x2.diag} are Hermitian  (to ensure
$\sigma_i=|\lambda_i|$), then we have $C_{l_p}(A)=C_{p}(A)$ for
all $p$ and a Hermitian $2\times 2$ matrix $A$. Since $\rho_i$'s are
effectively in $2\times 2$, we have
$C_{l_p}(\rho_i)=C_{p}(\rho_i)$. Similarly,
$C_{l_p}(\rho)=C_{p}(\rho)$, as $\rho$ is a matrix of the form
given in  \eeqref{Eq:Ctr(Qubit.Oplus.Scalar)}. Thus strong
monotonicity for $C_p$ is also violated for all $ab\neq 0$ and
$p\in(1,\infty)$.

Up to now we were concerned about only strong monotonicty of $C_p$ and $C_{l_p}$. 
It appears that for $p\in(1,\infty)$ none is a monotone in the first place.
\begin{proposition}
	\label{Prop:clp&cp.not.monotone} For $p\in(1,\infty)$, neither $C_p$ nor $C_{l_p}$ is a monotone.
\end{proposition}

Note that this result is stronger than Proposition~\ref{Prop:Clp&Cp.not.SM}, because, convexity together with strong monotonicity implies monotonicity. 
So, if a convex function is not a monotone, it can not be a strong monotone. 

It also appears that we can give a general method to construct counterexample from any coherent state \cite{Thanks.to.Alex}. Before doing so, we note that it follows from the result of \ccite{Garcia+3.JMP.2006}, that $C_p$ is a monotone for all qubit states and for all $p\in[1,\infty)$. So the counterexample should be in dimension higher than 2. 

The states themselves being incoherent, there is an incoherent channel transforming $\id/d$ to $|0\ran\lan0|$. For instance, consider the Kraus operators $K_i=|0\ran\lan i-1|$, $i=1,2,\dotsc,d$. Now let $\rho$ be a given coherent state and $\Lambda^I$ be the incoherent channel with Kraus operators $\tilde{K}_i=\id\otimes K_i$. Then we have \begin{align*}
C_p\left(\Lambda^I[\rho\otimes\id/d]\right)&=C_p\left(\rho\otimes|0\ran\lan 0|\right)\\
&=C_p\left(\rho\right)\\
&>C_p\left(\rho\otimes\id/d\right).
\end{align*}
In the last line we have used \begin{multline*}
C_p\left(\rho\otimes\id/d\right)\leq \norm{\rho\otimes\id/d-\delta^\star\otimes\id/d}_p=C_p(\rho)\norm{\id/d}_p<C_p(\rho).
\end{multline*}
Noticing that $C_{l_p}(\rho\otimes\id/d)=d^{1/p-1}C_{l_p}(\rho)<C_{l_p}(\rho)$, $C_{l_p}$ also violates monotonicity.

\section{\label{Sec:Discussion}Discussion and Conclusion}
Although originated in entanglement theory, strong monotonicity is
not a necessary requirement for entanglement measures, but rather
an extra feature. In contrast, every coherence measure has to
satisfy strong monotonicity. It would be interesting to study the
effect of relaxing this constraint. Also restricting the Kraus operators to have same dimension as that of the original state would be worth looking at.

The strong monotonicity of a convex entanglement measure is known
to be equivalent to its local unitary invariance and \emph{flag
condition} \cite{Horodecki.OSID.2005}. In
\ccite{Baumgratz+2.PRL.2014} a quite different flag condition has
been mentioned as an extra feature of a coherence measure. Since
trace norm is factorizable under tensor products, it follows that
if the strong monotonicity holds for $C_{\text{tr}}$, then it will
also satisfy the flag condition:
\[C_{\text{tr}}\left(\sum_ip_i\rho_i\otimes|i\ran\lan i|\right)\leq C_{\text{tr}}(\rho).\]
However, this does not help to resolve the main question, and
despite the frequent appearance of trace distance in literature,
it (at least for $E_{\text{tr}}$) remains quite a frustrating open
problem.

Before concluding, let us mention some relevance of our
Conjecture~\ref{Conj:cl1.geq.max}. As was mentioned earlier,
$C_{l_1}$ does not have any physical interpretation yet. In some
recent works \cite{Bera+3.PRA.2015,Bagan+2.Ar.2015}, $C_{l_1}$ has
been shown to be connected with the success probability of
unambiguous state discrimination in interference  experiments. If
the conjectured relation $C_{l_1}(\rho)\geq C_r(\rho)$ holds for
all states, then it would probably be the best physical
interpretation for $C_{l_1}$. It will then be analogous to
(logarithmic) negativity in entanglement theory, providing an
upper bound for distillable coherence (which coincides with
$C_r(\rho)$ for all $\rho$ \cite{Winter+Yang.Ar.2015}).
\begin{acknowledgments}
    We would like to thank Andreas Winter and Koenraad M. R. Audenaert for helpful discussion and correspondence. We also thank
    Remigiusz Augusiak, Alexander Streltsov, and Manabendra Nath Bera for insightful comments, and Shuming Cheng for bringing \ccite{Lin.IEEE.1991} to our attention. SR and ML acknowledge
    financial support from the John Templeton Foundation, the EU grants OSYRIS (ERC-2013-AdG Grant No. 339106), QUIC (H2020-FETPROACT-2014 No. 641122),
    and SIQS (FP7-ICT-2011-9 No. 600645), the Spanish MINECO grants FOQUS (FIS2013-46768-P) and ``Severo Ochoa" Programme (SEV-2015-0522), and the Generalitat de Catalunya
    grant 2014 SGR 874. 
\end{acknowledgments}

\end{document}